\documentclass{iaus}
\usepackage{graphicx}

\title[Massive Star and Star Cluster Formation] 
{Massive Star and Star Cluster Formation}

\author[Tan]   
{Jonathan C. Tan$^1$}

\affiliation{$^1$Department of Astronomy, University of Florida, Gainesville, FL 32611, USA
\break email: jt @ astro.ufl.edu}

\pubyear{2006}
\volume{237}  
\pagerange{???--???}
\date{?? and in revised form ??}
\jname{Triggered Star Formation in a Turbulent ISM}
\editors{B. G. Elmegreen \& J. Palous, eds.}
\begin{document}

\maketitle

\begin{abstract}
I review the status of massive star formation theories: accretion from
collapsing, massive, turbulent cores; competitive accretion; and
stellar collisions. I conclude the observational and theoretical
evidence favors the first of these models. I then discuss: the initial
conditions of star cluster formation as traced by infrared dark
clouds; the cluster formation timescale; and comparison of the initial
cluster mass function in different galactic environments.

\end{abstract}

\firstsection 
\section{Introduction}

Massive stars and star clusters form together as part of a single
unified process. All locally-observed massive stars appear to form in
star clusters (de Wit et al. 2005), particularly in rich star clusters
(Massi, Testi \& Vanzi 2006). Star clusters make a significant,
perhaps dominant, contribution to the total star formation rate of
galaxies (Lada \& Lada 2003; Fall, Chandar, \& Whitmore 2005), so to
understand global star formation properties of galaxies
(e.g. Kennicutt 1998), one must understand star cluster formation.

\section{Massive Star Formation}

There is still some debate about how massive stars form. Do they form
from the global collapse of a massive, initially starless gas core, in
which a central protostar or binary grows from low to high mass by
accretion from a disk (e.g. McKee \& Tan 2003)? This is a
scaled-up version of the standard model of low-mass star formation
(Shu, Adams, \& Lizano 1987). Or do they form from favored low-mass protostellar
seeds that accrete gas competitively, with the gas being bound to the
protocluster potential but not at any stage in a spatially coherent
bound core with a mass similar to that of the final massive star
(e.g. Bonnell, Vine, \& Bate 2004). These latter models typically involve the
global collapse of the protocluster gas over a timescale
approximately equal to its free-fall time, so the growth of the
massive star takes place on the same timescale as the formation of the
entire cluster. It has been suggested that protostellar collisions may
also be involved in the growth of massive stars (Bonnell, Bate, \&
Zinnecker 1998; Bally \& Zinnecker 2005).

Evidence in support of the core model of massive star formation
includes the fact that massive starless cores are observed and the
mass function of these cores is similar to the stellar initial mass
function (IMF) (Beuther \& Schilke 2004; Reid \& Wilson 2006).
Massive cores tend to have line widths that are much broader than
thermal (Caselli \& Myers 1995), indicating that other forms of
pressure support such as turbulent motions and magnetic fields are
important. Indeed observed magnetic field strengths are close to the
values needed to support the gas (Crutcher 2005).  Known massive
protostars tend to be embedded in dense gas cores with masses
comparable to the stellar masses (e.g. Source I in the Orion Hot Core;
W3($\rm H_2O$)). Low-mass protostars, i.e. actively accreting stars,
always have relatively massive accretion disks and outflows. A number
of claims have been made for disks around massive protostars, although
it is usually difficult to determine if these are rotationally
supported structures (see Cesaroni et al. 2006 for a review). Powerful
outflows from massive protostars with similar degrees of collimation
to those from low-mass protostars have been seen (Beuther et
al. 2002). The expected evolutionary scheme for high-mass star
formation from cores has been reviewed in more detail by Beuther et
al. (2006). Doty, van Dishoeck, \& Tan (2006) considered the chemical
evolution of this model with particular application to observations of
water abundance in hot cores. Kratter \& Matzner (2006) investigated
the gravitational stability of massive protostellar accretion
disks. Krumholz, McKee, \& Klein (2007) presented
radiation-hydrodynamic simulations of massive star formation from a
massive turbulent core.


Massive star formation models involving competitive accretion and
stellar collisions face several observational and theoretical hurdles.
Edgar \& Clarke (2004) showed that Bondi-Hoyle accretion becomes very
inefficient for protostellar masses $\gtrsim 10\:M_\odot$ because of
radiation pressure on dust in the gas. This feedback has not been
included in any of the simulations in which massive stars form by
competitive accretion.

To overcome radiation pressure the accretion flow to a massive star
must become optically thick, either in a dense core or disk, or in
collisions of protostars.  The collisional timescale is $t_{\rm
coll}=1.44\times 10^{10} (n_*/10^4{\rm pc^{-3}})^{-1} (\sigma/2{\rm
km\:s^{-1}}) (r_*/10R_\odot)^{-1} (m_*/M_\odot)^{-1}\:{\rm yr}$ in the
limit of strong gravitational focusing, where $\sigma$ is the 1D
velocity dispersion and $r_*$ is the radius of the stellar collisional
cross-section. For collisions to occur frequently enough to grow a
massive protostar within $10^6\:{\rm yr}$ (massive zero age main sequence
stars are observed) requires protostellar densities of at least
$10^6\:{\rm pc^{-3}}$ and probably closer to $10^8\:{\rm pc^{-3}}$,
whereas typical observed stellar densities around massive protostars
are much smaller. For example from the Orion Nebula Cluster (ONC) x-ray
observations of Garmire et al. (2000), Tan (2004) estimates a stellar
density of about $10^5\:{\rm pc^{-3}}$ in the KL region. This result
is not significantly changed by the deeper x-ray observations of
Grosso et al. (2005). Hunter et al. (2006) find a density of sub-mm
cores in the center of protoclusters in NGC~6334 of about $10^4\:{\rm
pc^{-3}}$. From Fig.~1 and the data of Mueller et al. (2002) we see
that typical mean densities of the central regions of Galactic
protoclusters are $n_{\rm H}\simeq 2 \times 10^5\:{\rm
cm^{-3}}$, i.e. 7000~$M_\odot\:{\rm pc^{-3}}$. If all this
gas mass formed stars, stellar
densities would be about $10^4\:{\rm pc^{-3}}$, given a typical
IMF. The fiducial core that forms a massive star in the model of McKee
\& Tan (2003) is also shown in Fig.~1, and has a mean density about
one ten times greater than this. Even if the core fragmented
with 100\% efficiency into low-mass stars the stellar density would be
too low for efficient growth via stellar collisions. In fact
numerical simulations show that fragmentation of the core into many
stars is impeded by heat input from the forming central massive
star (Krumholz 2006). The numerical simulations in which greater
degrees of fragmentation are seen (e.g. Dobbs, Bonnell, \& Clark 2005)
do not include this feedback. Magnetic pressure is also likely to be
important for the support of cores more massive than the thermal Jeans
mass, but this is also usually not included in simulations of massive
star formation.

If collisions are relevant for massive star formation, but not
low-mass star formation, then one might expect a
change in the slope of the stellar IMF at the
mass scale at which the collisional process becomes important. In fact
the stellar IMF is reasonably well-fit by a power law from $\sim 1\:M_\odot$ out to the
highest observed masses (Massey 1998).

\begin{figure}
\includegraphics[height=5.5in,width=5.5in,angle=0]{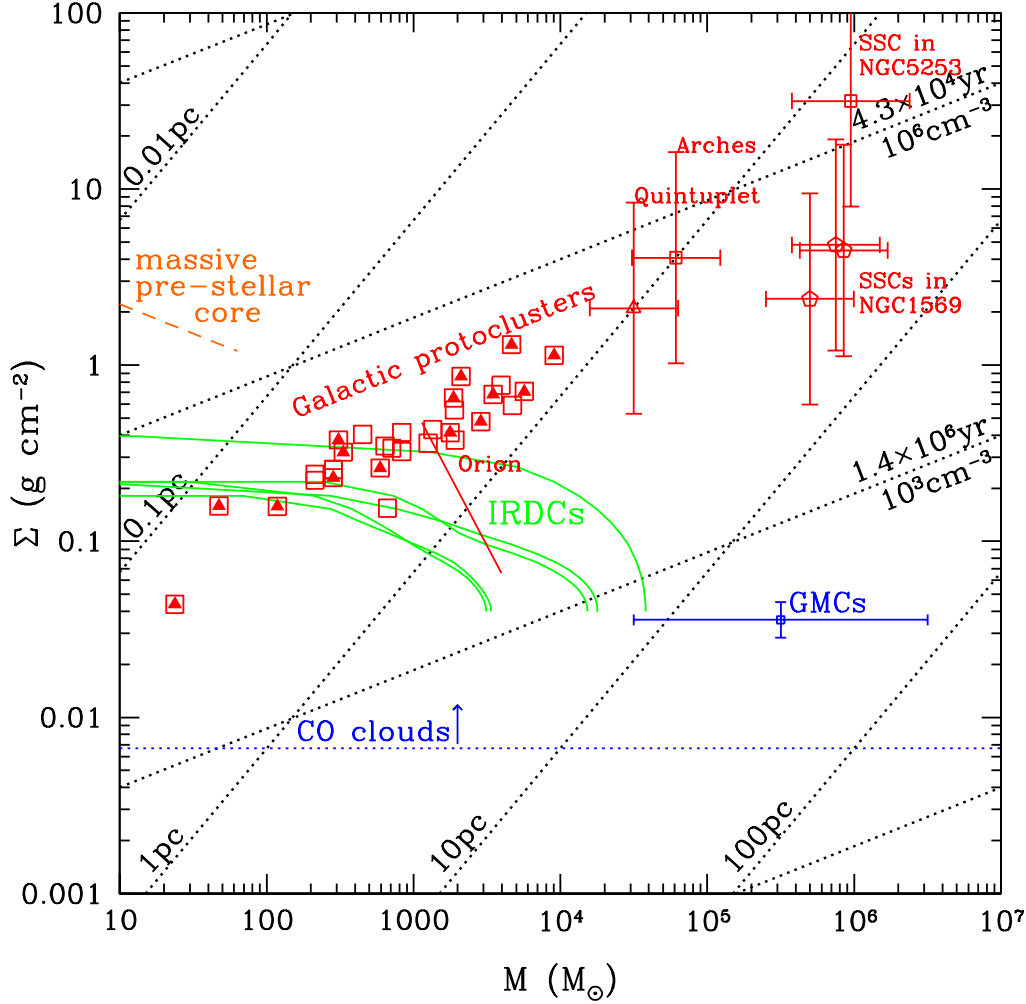}
  \caption{Surface density, $\Sigma\equiv M/(\pi R^2)$, versus mass,
  $M$, for star clusters and interstellar clouds. Contours of constant
  radius, $R$, and hydrogen number density, $n_{\rm H}$, or free-fall
  timescale, $t_{\rm ff}$, are shown with dotted lines. The minimum
  $\Sigma$ for CO clouds in the local Galactic FUV radiation field is
  shown, as are typical GMC parameters and the distributions
  $M(>\Sigma)$ of several IRDCs derived from extinction mapping
  (Butler et al., in prep.). Open squares are star-forming clumps
  (Mueller et al. 2002): a triangle indicates presence of an HII
  region.  The solid straight line traces conditions from the inner to
  outer parts of the ONC, assuming equal mass in gas
  and stars. Several more massive clusters are also indicated. The
  fiducial massive core in the model of McKee \& Tan (2003) is shown
  by the dashed line.
\label{fig:1}}
\end{figure}

\section{Star Cluster Formation}

\subsection{The Initial Conditions for Star Cluster Formation: Infrared Dark Clouds}

We expect the initial conditions for star clusters to be the densest
starless gas clouds. Such clouds reveal themselves by absorption of
the Galactic diffuse infrared background and have become known as
Infrared Dark Clouds (IRDCs) (Egan et al. 1998).

\begin{figure}
\includegraphics[height=2.65in,width=2.65in,angle=0]{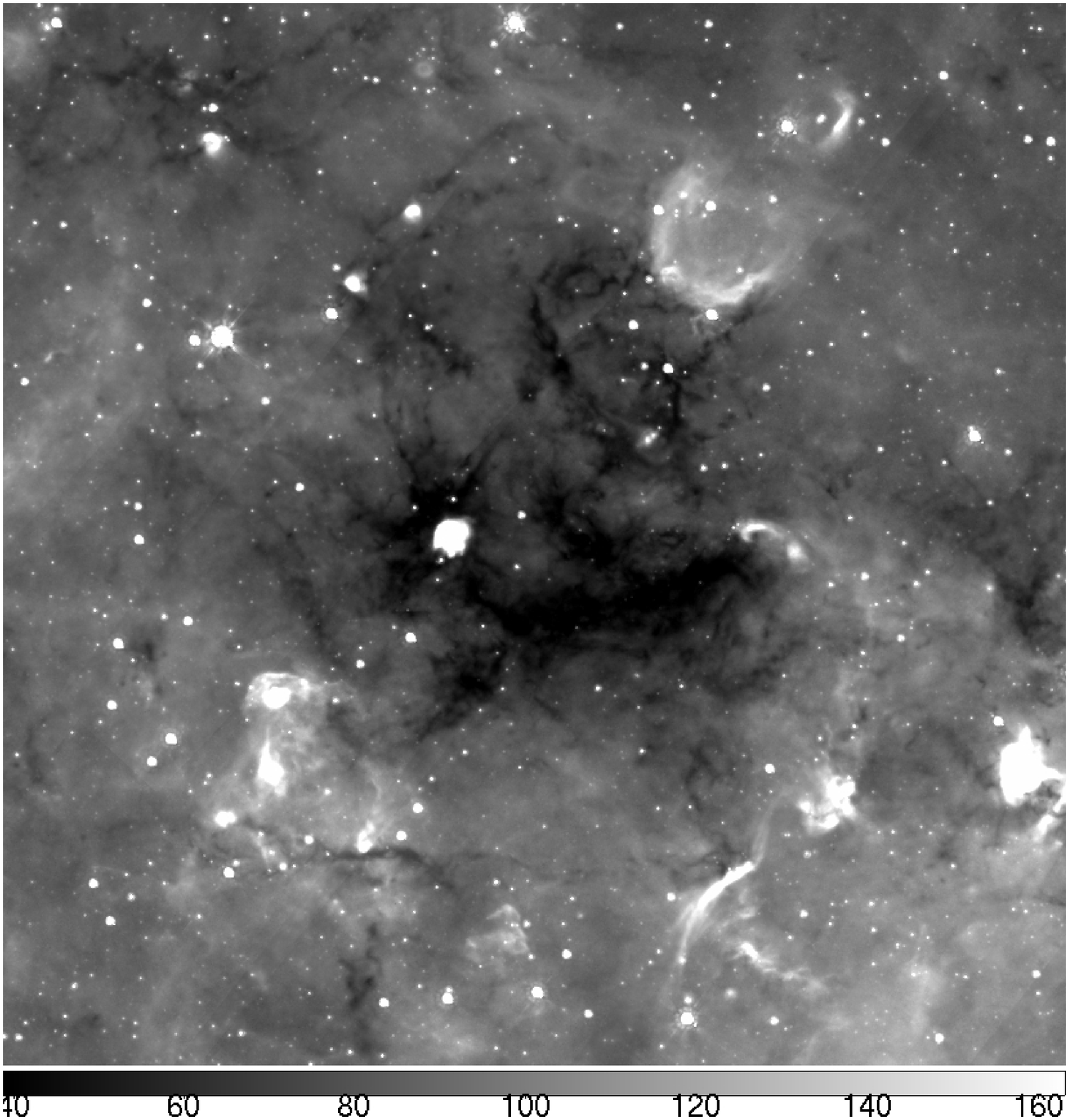}
\includegraphics[height=2.65in,width=2.65in,angle=0]{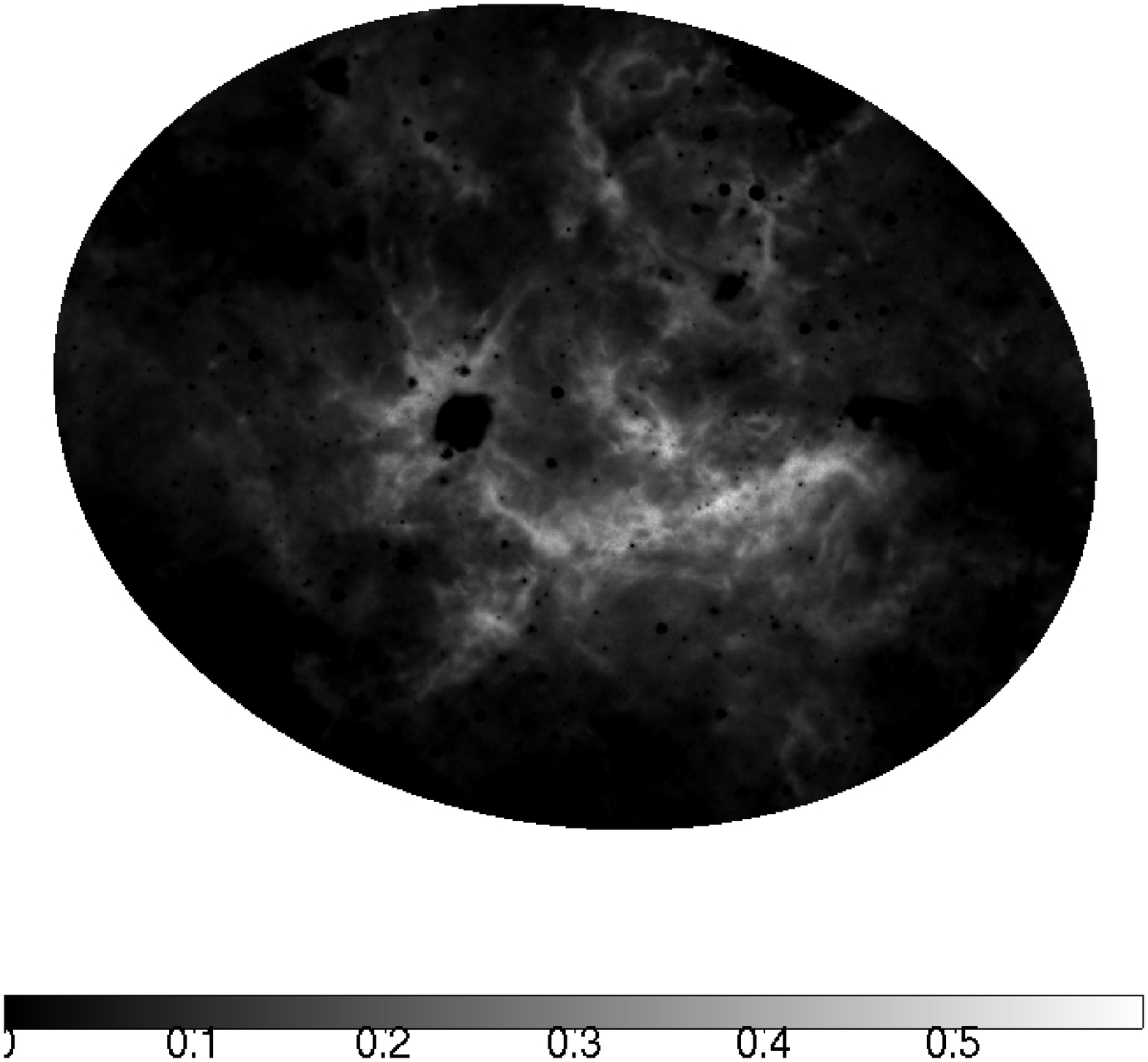}
  \caption{(a) Left: Example IRDC at $l=28.37$, $b=0.07$ and distance
  4.9~kpc observed by Spitzer at 8~$\rm \mu m$. Image is $16.5^\prime$ across. (b) Right: $\Sigma$
  map of the same cloud derived from extinction of the diffuse
  background (Butler et al., in prep). Intensity scale is in
  $\rm g\:cm^{-2}$. Note the extinction mapping technique
  fails where there is a bright source in front of or in
  the cloud.}
\label{fig:irdc}
\end{figure}

One way to measure the physical properties of these clouds is through
extinction mapping (Fig.~\ref{fig:irdc}). Assuming the diffuse
Galactic infrared emission behind the cloud is similar to that around it
and adopting an infrared extinction law and dust to gas
ratio (Weingartner \& Draine 2001) allows the measurement of mass
surface density, $\Sigma$. A kinematic distance can be
measured from $\rm^{13}CO$ line emission (Simon et al. 2001), and thus
the physical size and mass of the cloud determined. The cummulative
distributions of $M(>\Sigma)$, i.e. the mass that is at surface
densities greater than or equal to a particular $\Sigma$, for five
typical IRDCs have been measured by Butler, Tan, \& Hernandez, in prep
and are shown in Fig.~1. The IRDCs span physical properties
similar to those of embedded star clusters (Mueller et al. 2002), although with somewhat lower surface densities, and
so are likely to be representative of the initial conditions of star
cluster formation. 


\subsection{The Timescale for Star Cluster Formation}

Some models of massive star and star cluster formation involve the
global collapse of the protocluster in about one free-fall time
(e.g. Bonnell, Vine \& Bate 2004), while other models that include
feedback from the forming stars (e.g. Li \& Nakamura 2006) have star
formation occuring more gradually over at least several free-fall
times. Based on the results of numerical simulations, Krumholz \&
McKee (2005) argued that supersonically turbulent gas forms stars at a
slow rate of only a few percent of the total gas mass per dynamical or
free-fall time. Tan, Krumholz, \& McKee (2006) extended this analysis
to spherical clumps and argued that those clumps that
eventually turn a high ($\gtrsim30\%$) fraction of their mass into
stars must do so over at least several ($\gtrsim 7$) free-fall times.

Tan et al. (2006) also summarized the observational evidence in
support of slow, quasi-equilibrium star cluster formation: (1) The
morphologies of CS gas clumps are round (Shirley et al. 2003); (2) the
spatial distributions of stars in embedded, rich, i.e.
high-star-formation-efficiency, star clusters show relatively little
substructure; (3) the momentum flux from the combined outflows from
protostars in forming clusters is relatively small; (4) the age
spreads of stars in rich star clusters are much greater than their
current free-fall times; (5) in the ONC a dynamical
ejection event associated with the cluster has been dated at 2.5~Myr
(Hoogerwerf, de Bruijne, \& de Zeeuw 2001), which is much longer than
the free-fall time of the present cluster.

Note it is the central, high-star-formation-efficiency region of the
cluster where we propose that star formation takes place over several
to many free-fall times. These regions have short free-fall times (see
Fig.~1), $\sim 10^5\:{\rm yr}$. The outskirts of the cluster have much
lower densities, longer free-fall times, and star formation here may
occur over just one or two free-fall times, as proposed by Elmegreen
(2000), before gas is disrupted by feedback from the newly-formed
cluster. The global star formation efficiency here will be relatively
low and the young stars will exhibit a greater degree
of substructure.

It has been suggested that the star cluster formation process takes
only one or two free-fall times when this time is referenced to the
pre-cluster conditions at lower density, and that therefore star
cluster formation can be regarded as being the result of dynamic
collapse of a cloud and is not a quasi-equilibrium process (Hartmann \&
Burkert 2006). This distinction is important because Krumholz, McKee,
\& Klein (2005) showed that the process of star formation by
competitive accretion cannot be important in virialized, equilibrium
clouds. It requires sub-virial conditions associated with global
collapse. 
Several arguments can be made against the global collapse picture: (1)
the protocluster gas clouds appear to be approximately virialized
(e.g. Shirley et al. 2003); (2) the final distributions of the
distances of the stars from the cluster center should reflect the
locations at which they formed or be even larger because of gas
removal, yet we see newly formed rich star clusters with concentrated,
dynamically-relaxed distributions. Huff \& Stahler (2006) found the
star formation history of the ONC showed no dependence on the radial
distance from the cluster center; (3) again in the ONC, the 2.5~Myr
dynamical ejection event (Hoogerwerf et al. 2001) suggests that at
this time the cluster was already in a state of high stellar density
with a short free-fall time.


\subsection{The Initial Cluster Mass Function}

The initial cluster mass function (ICMF) is a fundamental property of
the star cluster formation process. If there are external triggers,
e.g. cloud collisions, supernova blast waves, that initiate star
cluster formation, then these may influence the ICMF.  It has been
suggested that super star cluster formation may be favored in the
low-shear environment of dwarf irregulars (Billett, Hunter, \&
Elmegreen 2002).

To investigate whether the ICMF depends on galactic environment,
Dowell, Buckalew, \& Tan (2007) used automated source selection from
Sloan Digital Sky Survey (SDSS) data to measured the ICMF at masses
$\gtrsim 3\times 10^4\:M_\odot$ in 13 nearby ($\lesssim10$~Mpc) dwarf
irregular galaxies, which tend to have relatively low metallicity and
shear. Cluster ages, masses and reddening were determined by comparing
Starburst99 models with the multi-color photometry. Completeness
corrections were made, although these are relatively small for massive
young clusters at these distances. Foreground stellar and background
galactic contamination were assessed and found to be small. The ICMF
was assumed to be equal to the mass function of clusters with ages
$\leq 20$~Myr. The same procedure was repeated on SDSS data of several
nearby spiral galaxies at similar distances but with higher metallicity
and shear. Several hundred clusters were identified from both the
dwarf irregular and spiral galaxy samples.

The main result is that these samples are stastically
indistinguishable from each other, suggesting that the ICMF does not
depend on galactic shear or metallicity. We find the ICMF is
reasonably well fit by a power law $\frac{dN(M)}{dM} \varpropto
M^{-\alpha_M}$ with $\alpha_M \simeq 1.5$ in both dwarf irregular and
spiral galaxies. This is somewhat shallower than the power law index
of $\alpha_M\simeq2$ that has been found in spiral galaxies by
(Larsen 2002) using HST images. This may be due to the lower resolution
of the SDSS observations, which lead to blending of clusters that form
within $\sim 50$~pc of each other. Nevertheless the similarity of the
cluster (or association) mass functions between the galaxy samples
suggests that a universal process, perhaps turbulent fragmentation
inside GMCs (Elmegreen \& Efremov 1997), is responsible for star
cluster formation.

\begin{acknowledgments}
JCT acknowledges support from CLAS, University of Florida.
\end{acknowledgments}

\begin{discussion}

\discuss{Linz}{The speaker mentioned that apparently no current
star formation occurs in his sample of IRDCs. Can we be really
sure about that, since in most cases, objects are found in IRDCs
by Spitzer/MIPS?}

\discuss{Tan}{There are two issues here: (1) at each location in the
cloud there is a constraint on the embedded luminosity from the lack
of flux at $\sim 8\:{\rm \mu m}$, and, without having done detailed
calculations, my impression is that for most of the regions of IRDCs
in our sample there is no current, active, luminous star formation,
i.e. massive star formation, occurring. There could be embedded
lower-luminosity sources and it would be useful to probe this
population (either with Spitzer/MIPs or with x-rays). (2) IRDCs are
not a particularly well-defined class of objects, and there can in
fact be bright sources nearby in adjacent clouds or even in part of
the same cloud (the cloud in Fig. 2 has such a source). Still, if one
were to measure the total light to mass ratios of IRDCs these should on
the average be quite low compared to more evolved star-forming clouds.}

\discuss{Linz}{Still, these objects embedded there might be lower
luminosity now but could develop into high-mass YSOs later on?}

\discuss{Tan}{I agree that many or most IRDCs, especially the
relatively high column density ones that we are studying, are likely
to form star clusters and massive stars in the future.}

\discuss{Fukui}{In your turbulent picture, how could you explain the
formation of super star clusters?}

\discuss{Tan}{Observed super star clusters (SSCs) have $\sim
10^6\:M_\odot$ inside a sphere of radius $\sim 3~{\rm pc}$. One basic open
question is whether the initial condition is an essentially starless
gas cloud with these properties or whether SSCs form more gradually as
smaller clouds (perhaps already forming star clusters) merge with the
main cluster. In my opinion, it would be difficult to produce the
starless initial condition from typical Galactic GMCs without some
kind of synchronized, fast trigger. The escape speed from SSCs is
greater than the ionized gas sound speed, so they may be forming with
very high efficiency from their parent gas clouds (requiring long
formation times [and age spreads] in terms of free-fall times) (Tan \&
McKee 2004, in proc. of Cancun Workshop). This longer formation time
may allow more time for infall and merger of surrounding gas clouds,
and the higher efficiency means less total gas mass is needed to reach
the final stellar mass.}

%

\end{discussion}

\end{document}